\documentclass[12pt]{article}

\usepackage{amssymb}
\usepackage{amsmath,graphicx,amsfonts,epsfig,eucal}
\usepackage{amsfonts}
\usepackage{color}
\usepackage{amsthm}

\usepackage[utf8]{inputenc}
\usepackage[english]{babel}

\newtheorem{definition}{Definition}

\begin{document}
\title{Closed-form solutions of Lucas–Uzawa model with externalities via partial Hamiltonian approach. Some Clarifications}
\author{Constantin Chilarescu}
\date{}

\maketitle

\centerline{\it Laboratoire CLERSE Universit\'{e} de Lille, France}

\centerline{\it E-mail: Constantin.Chilarescu@univ-lille.fr}

\maketitle

\begin{abstract}
The main aim of this paper is to give some clarifications to the recent paper published in Computational and Applied Mathematics by Naz and Chaudhry.
\end{abstract}

{\small {\bf Keywords}: Lucas–Uzawa model with externalities; Uniqueness of solutions;

\hspace{2.05cm} Partial Hamiltonian approach.}



\date{}
\maketitle

\section{The Model of Lucas with externality}
The two sectors model considered in the paper of Lucas $(1988)$, in terms of per capita quantities, is given by the following definition.
\begin{definition}
The set of paths $\left\{k, h, c, u\right\}$ is called an optimal
solution if it solves the following optimization problem:
\begin {equation}\label{eqP}
V_0 = \max\limits_{u,c}\int\limits_0^{\infty}\frac{\left[c(t)\right]^{1-\sigma}-1}{1-\sigma}e^{-\rho t}dt,
\end {equation}
\noindent subject to
\begin {equation}\label{eqres}
\left\{
 \begin{array}{lll}
\dot{k}(t) = \gamma\left[k(t)\right]^\beta\left[h(t)\right]^{1-\beta+\theta}\left[u(t)\right]^{1-\beta} - \pi k(t) - c(t),\\\\
\dot{h}(t) = \delta[1 - u(t)]h(t),\\\\
k_0 = k(0),\;h_0 = h(0),
  \end{array}
  \right.
\end {equation}
\noindent where $k$ is physical capital,
$h$ is human capital, $c$ is the real per-capita consumption and $u$
is the fraction of labor allocated to the production of physical capital, with $k_0 > 0$ and $h_0 > 0$ being given. $\beta$ is the
elasticity of output with respect to physical capital, $\rho$ is a
positive discount factor, the efficiency parameters $\gamma > 0$ and
$\delta > 0$ represent the constant technological levels in the good
sector and, respectively in the education sector $\theta$ is a positive externality parameter and
$\sigma^{-1}$ represents the constant elasticity of intertemporal
substitution, and throughout this paper we suppose that $\sigma \neq 1$ and $\sigma \neq \beta$.
\end{definition}
\noindent The equations $\eqref{eqres}$ give the resources constraints and initial
values for the state variables $k$ and $h$. Of course, the two state variables and the two control variables $c$ and $u$ are all functions of times, but when no confusions are possible, we simply write $k, h$, $c$ and $u$.
To solve the problem
$\eqref{eqP}$ subject to $\eqref{eqres}$, we define the Hamiltonian function:
$$
H = \frac{c^{1-\sigma}-1}{1-\sigma} +
\left[\gamma k^\beta\left(h^{\frac{1-\beta+\theta}{1-\beta}}u\right)^{1-\beta} - \pi k - c
\right]\lambda + \delta(1 - u)h\mu.
$$
\noindent The boundary conditions include initial values $(k_0,
h_0)$, and the transversality conditions:
$$\lim\limits_{t\rightarrow\infty} e^{-\rho t}\lambda(t)
k(t) = 0 \;\;\mbox{and}\;\; \lim\limits_{t\rightarrow\infty}
e^{-\rho t}\mu(t) h(t) = 0.$$
Differentiating the Hamiltonian with respect to $c$, $u$, $k$ and $h$, we obtain the following dynamical system that drives the economy over time.
\begin {equation}\label{eqfoc}
\left\{
  \begin{array}{llllll}
\dot{k} = \left[\gamma\left(\frac{h^{\frac{1-\beta+\theta}{1-\beta}}u}{k}\right)^{1-\beta} - \pi\right]k
- c,\\\\
\dot{h} = \delta[1 - u]h,\\\\
\dot{c} = \left[-\frac{\rho+\pi}{\sigma}+\frac{\gamma\beta}{\sigma}\left(\frac{h^{\frac{1-\beta+\theta}{1-\beta}}u}{k}\right)^{1-\beta}\right]c,\\\\
\dot{u} = \left[\frac{(\delta+\pi)(1-\beta)+\theta\delta}{\beta}-\frac{c}{k}+\frac{\delta(1-\beta+\theta)}{1-\beta}u\right]u,\\\\
\dot{\lambda} = \left[\rho+\pi-\gamma\beta\left(\frac{h^{\frac{1-\beta+\theta}{1-\beta}}u}{k}\right)^{1-\beta}\right]\lambda\\\\
\dot{\mu} = \left[\rho-\delta-\frac{\theta\delta}{1-\beta}u\right]\mu.\\
\end{array}
  \right.
\end {equation}
The alternative of the above model, obtained via Hiraguchi transform, was analyzed, first of all by Boucekkine and Ruiz-Tamarit $(2008)$, and later by Chilarescu $(2011)$ and both papers proved, doubtless that the model possesses a unique solution. More clarifications on the uniqueness of solutions to the model of Lucas can be found in a recent paper of Chilarescu $(2018a)$ .

The model of Lucas with externalities was studied by Hiraguchi $(2009)$ and he proved in his paper that this model possesses a unique set of solutions. The method employed by Hiraguchi was that of hypergeometric functions. In a recent paper of Chilarescu $(2018b)$, the same model of Lucas with externalities was completely solved, this time in a simpler manner, using only classical mathematical tools. He also provided a proof of the existence and uniqueness of solutions.

In the paper we comment, Naz and Chaudhry $(2018)$ claim that they found two different set of solutions for the model of Lucas with externalities. The first set of solutions coincides exactly with those determine by Hiraguchi and later by Chilarescu. In the second set, the solutions for the variables $k$ and $c$ are identical with those of the first set, that is:
\begin{equation}\label{solk}
k(t) = \frac{k_0z_0}{F_*}\left[z(t)\right]^{-1}\left[F_*-F(t)\right]e^{\phi t},\;\phi=\frac{(1-\beta)\left[\delta+\pi(1-\beta)\right]+\theta\delta}{\beta(1-\beta)},
\end{equation}
\begin{equation}\label{solc}
c(t) = \frac{k_0z_0}{F_*}\left[z(t)\right]^{-\frac{\beta}{\sigma}}e^{\chi t},\;\chi=\frac{(1-\beta)(\delta-\rho)
+\theta\delta}{\sigma(1-\beta)},
\end{equation}
where
$$z(t) = \frac{\left[h(t)\right]^{\frac{1-\beta+\theta}{1-\beta}}u(t)}{k(t)},\;F(t) = \int\limits_0^t z(s)^{\frac{\sigma-\beta}{\sigma}}e^{-\xi s}ds,$$ $$\xi=\phi-\chi,\;F_*=F_*(u_0) = \lim\limits_{t\rightarrow\infty}F(t),$$
whereas the solutions for the variables $h$ and $u$ are different. In the first set the solutions they found are:
\begin{equation}\label{solhs1}
h(t) = h_0\left\{\frac{u_0e^{\phi t}\left[F_*-F(t)\right]}{F_*u(t)}\right\}^{\frac{1-\beta}{1-\beta+\gamma}},
\end{equation}
\begin{equation}\label{solus1}
u(t) = \frac{\varphi u_0\left(F_*-F(t)\right)}{\left[\left(\varphi+\delta\eta u_0\right)F_*-\delta\eta u_0B(t)\right]e^{-\varphi t}-\delta\eta u_0\left[F_*-F(t)\right]},
\end{equation}
where
$$\eta=\frac{1-\beta+\theta}{1-\beta},\;\varphi=\frac{(\delta+\pi)(1-\beta)+\gamma\delta}{\beta},$$ $$B(t) = \int\limits_0^t z(s)^{\frac{\sigma-\beta}{\sigma}}e^{-(\xi-\varphi) s}ds,
\,B_*=B_*(u_0) = \lim\limits_{t\rightarrow\infty}B(t)=\left(1+\frac{\varphi}{\delta\eta u_0}\right)F_*.$$
The corresponding solutions determined in the second set are:
\begin{equation}\label{solhs2}
h(t) = h_0\left\{\frac{u_0e^{\phi t}\left[F_*-F(t)\right]}{F_*u(t)}\right\}^{\frac{1-\beta}{1-\beta+\gamma}},
\end{equation}
but with $u(t)$ given by
\begin{equation}\label{solus2}
u(t) = \frac{u_0}{k_0}\frac{\left\{z_0^{\beta-1}\left[\sigma c_0-\left(\rho+\pi-\pi\sigma\right)k_0\right]+\gamma\beta(1-\sigma)k_0\right\}\left[F_*-F(t)\right]}{
\left[\gamma\beta(1-\sigma)-\left(\rho+\pi-\pi\sigma\right)z^{\beta-1}\right]\left[F_*-F(t)\right]+\sigma z^{\beta-\frac{\beta}{\sigma}}e^{-\xi t}}.
\end{equation}
At this point we have the following comments.
\begin{enumerate}
  \item [1.] Because the solutions for $k$ and $c$ are the same in both sets of solutions, we can substitute these results into the fourth equation of the system \eqref{eqfoc}, to obtain:
      \begin{equation}\label{equt}
      \dot{u} = \left[\varphi - \frac{z^{\frac{\sigma-\beta}{\sigma}}e^{-\xi t}}{F_*-F(t)}
+\delta\eta u\right]u.
      \end{equation}
    As was proved by Chilarescu $(2018b)$, the starting value $u_0$ can be determined and is the unique solution of the equation
$$\left(\varphi +\delta\eta u_0\right) F_*(u_0) - \delta\eta u_0 B_*(u_0) = 0.$$
Consequently, since the function
$$F(t,u) = \left[\varphi - \frac{z^{\frac{\sigma-\beta}{\sigma}}e^{-\xi t}}{F_*-F(t)}
+\delta\eta u\right]u,$$
is continuously differentiable, than via the existence and uniqueness theorem for nonlinear differential equations, there exists one and only one solution to the initial value problem
$\dot{u} = F(t,u), \;u_0=u(0)$ and this solution is given by \eqref{solus1}.
  \item [2.] The authors only claim that their new solution for the control variable $u$ is an admissible solution, that is $u\in (0, 1)$ but they provided no proof. The proof is absolutely necessary.
  \item [3.] If this solution really exists, then the authors would have provided the proof that this results is completely different from that one produced by Hiraguchi and Chilarescu. At least they would have been able to supply graphs showing that the two trajectories are totally different. None of these requirements could be found in the paper of Naz and Chaudhry.
  \item [4] In our opinion, the so-called new solution determined by Naz and Chaudhry, is nothing else, than the same solution provided by Hiraguchi and Chilarescu, but only written in a different mathematical formulation.
\end{enumerate}

\end{document}